# A 3-dimensional interdigitated electrode geometry for the enhancement of charge collection efficiency in diamond detectors

J. Forneris, A. Lo Giudice, P. Olivero, F. Picollo, A. Re
*Physics Department and "NIS" Inter-departmental centre, University of Torino; Istituto Nazionale di Fisica Nucleare (INFN), Sezione di Torino; via P. Giuria 1, 10125 Torino, Italia*

Marco Marinelli, F. Pompili, C. Verona, G. Verona Rinati
*"Tor Vergata" University, Industrial Engeneering Department, via del Politecnico 1, 0133, Roma Italia*

M. Benetti, D. Cannata, F. Di Pietrantonio
*"O.M.Corbino" Institute of Acoustics and Sensors , CNR, Via del Fosso del Cavaliere 100, 00133 Roma, Italia.*

**Abstract** – In this work, a single crystal CVD diamond film with a novel three-dimensional (3D) interdigitated electrode geometry has been fabricated with the Reactive Ion Etching (RIE) technique in order to increase the charge collection efficiency (CCE) with respect to that obtained by standard superficial electrodes. The geometrical arrangement of the electric field lines due to the 3D patterning of the electrodes results in a shorter travel path for the excess charge carriers, thus contributing to a more efficient charge collection mechanism. The CCE of the device was mapped by means of the Ion Beam Induced Charge (IBIC) technique. A 1 MeV proton micro-beam was raster scanned over the active area of the diamond detector under different bias voltage conditions, enabling to probe the charge transport properties of the detector up to a depth of 8 μm below the sample surface. The experimental results, supported by the numerical simulations, show a significant improvement in the 3D-detector performance (i.e. CCE, energy resolution, extension of the active area) if compared with the results obtained by standard surface metallic electrodes.

**Introduction.** – Diamond has extreme physical and electronic properties such as high thermal conductivity, wide band gap, high breakdown field and high carrier mobilities [1, 2]. These properties make diamond an ideal candidate for the fabrication of radiation sensors especially for photon and particle detection where other materials hardly reach good performances. It is well known that the electronic properties of diamond can be influenced by recombination and trapping of free carriers on defect-induced localized states in the band-gap that may results in inadequate signal response and stability [3, 4]. However, the chemical vapour deposition (CVD) offers a technology for producing high-quality, intrinsic homoepitaxially grown layers on low-cost single crystal diamond substrates under tightly controlled growth conditions [5,6].

To this purpose, detectors based on high-quality homoepitaxially grown CVD single-crystal diamond were developed at the Department of Industrial Engineering of University of Rome "Tor Vergata" for both nuclear particles and photon radiation [7, 8].

Different device structures including photoresistive devices and Schottky photodiode both in sandwich or planar interdigitated structure have been reported in literature [9, 10]. Many authors have reported investigations of single and polycrystalline CVD diamond detectors with planar metal electrodes placed on the top of diamond surface [11-13,A]. However, due to the detection geometry of the device, producing an inhomogeneous electric field substantially confined in the region close to the detector surface, they show a low charge collection efficiency especially for ion detection[13, 14, A]. For this reason, recently new charge collection geometries have been explored through the fabrication of 3D graphitic electrodes in the diamond bulk, either by ion- [15] or laser-microbeam [16-18] direct writing.

In this paper, the fabrication, modeling and characterization of a novel three-dimensional interdigitated electrode geometry on single crystal CVD diamond layer is presented and studied. The 3D device has been developed in order to improve its charge collection efficiency (CCE). In particular, the detector was fabricated by hard metal masking followed by Reactive Ion Etching (RIE). Its CCE response was compared with the response of a similar detector based on standard superficial electrodes by means of Ion Beam Induced Charge (IBIC) mapping. Finally, the CCE performance was numerically modeled by means of Finite Element Methods (FEM) codes.

**Experimental** – *Sample preparation.* The fabrication process of the novel detector is shown in Fig. 1. A (50 ± XX)

µm thick intrinsic diamond layer was grown by Microwave Plasma Enhanced Chemical Vapour Deposition (MPECVD) technique on a commercial low-cost HPHT diamond substrate (Fig. 1a). The characterization of the deposited material in terms of optical spectroscopy and detection properties were reported in references x and y.

The diamond layer was oxidized after growth by isothermal annealing at 500 °C for 1h in air [19], in order to remove the H2 surface conductive layer. A 1 µm thick nickel film was then deposited by sputtering on the CVD diamond surface and subsequently a Ni mask with "interdigitated finger" shape was fabricated by photolithography (Fig. 1b). In order to etch the CVD diamond film, an anisotropic reactive ion etching (RIE) was performed (Fig. 1c). A gas mixture composed of 25% of CF4 in O2 was fluxed in the vacuum chamber with a flow rate of 20 sccm, reaching a pressure of 20 mbar; an etching rate of 3.8 µm h$^{-1}$ was obtained with a RF power of 200 W. At the end of the process, a groove depth of about 6 µm was achieved. A SEM micrograph of the interdigitated finger grooves in diamond is shown in Fig. 1d.

The Ni mask was then removed by chemical etching from the diamond surface (Fig. 1e). The interdigitated metallic electrodes were fabricated using a further lithographic step: a 50 nm thick chromium layer was firstly deposited on the diamond surface by thermal evaporation, then, with an aligned-lithography process, the Cr was selectively removed from the surface, leaving the metal only inside the grooves (Figs. 1f, 1g). The resulting structure consists of two separate combs of metal finger arrays of the 6 µm deep grooves, forming 10 µm wide interdigitated electrodes with a spacing of 10 µm. Such diamond detector will be referred to as "3D detector" in the following text (Fig.1h).

A second device was fabricated by depositing a nominally identical diamond layer. After CVD diamond growth, standard surface metallic contacts were patterned by a lift-off photolithographic technique and by thermal evaporation of a 50 nm thick Cr layer on the diamond surface. Both the width and the gap between the electrodes were 10 µm, in order to provide an equivalent sample for direct comparison with the performances of the 3D detector. In the following, this diamond detector will be referred to as "planar detector" (Fig.1i).

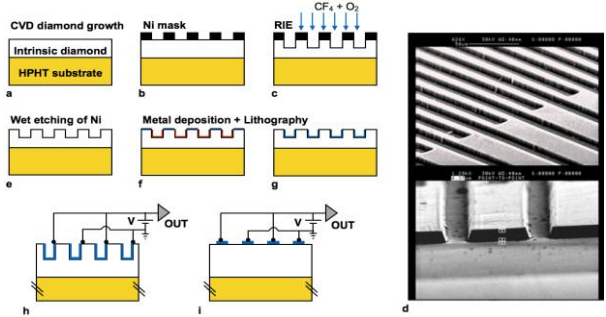

Fig. 1: (Colour on-line) Sketch of the fabrication process of the 3D detector: a) MPECVD growth of an intrinsic diamond layer over a HPHT substrate; b) deposition of Ni patterned mask c); diamond etching by in CF4+O2 plasma; d) SEM micrograph of the etched structures in the 3D detector; e) wet etching of the Ni mask; f) deposition of Cr contacts and g) removal of the Cr layer outside of the grooves; sketch of the electrical connections of h) the 3D detector and i) the planar detector.

In both devices, the metal contacts form a Schottky barrier with the CVD diamond surface [20, 21], which has a typical effective acceptor concentration of the order of ~$10^{14}$ cm$^{-3}$, as reported in previous works [22, Ciancaglioni]. An I-V characteristic (not reported here) of the 3D detector showed a linear, symmetric behavior, reaching typical values of 1 pA at ±20 V. Due to the planar geometry of the Cr grooves, the observed trend was attributed to the reverse current associated with the presence of back-to-back Schottky contacts at the electrodes, preventing a direct polarization current flowing through the device.

*IBIC experimental setup*. Both detectors were characterized by means of the Ion Beam Induced Charge (IBIC) [23] technique in order to evaluate their ionizing radiation detection performance. IBIC measurements were carried out at the micro-beam facility of the Italian INFN Legnaro National Laboratories (AN2000 accelerator) by raster-scanning a rarefied 1 MeV proton micro-beam focused to about 7 µm diameter spot, over rectangular areas (170×25 µm$^2$) of the active region of the diamond detectors. The energy of the ions was chosen to probe regions located beneath the metallic electrodes at 8 µm in depth [24], which is a suitable range to investigate the enhancement of the charge collection efficiency (CCE) in the 3D structure. A schematic representation of the electrical connections is shown in Figs. 1h and 1i: the bias voltage was applied to one of the two frontal electrode combs, whereas the other comb was grounded.

Measurements were performed at low ion current (~100 ions s-1) in order to avoid radiation damage and pile-up effects. The IBIC pulses were acquired by an electronic chain consisting of a charge-sensitive preamplifier ORTEC142A and an ORTEC572 shaping amplifier (gain 500, shaping time 0.5 µs). Pulse height processing, beam scanning and 2D map acquisition were performed using the commercial OMDAQ hardware and software system. The calibration of the electronic chain was performed using a Si detector (AMETEK TU-014-100-300) and a precision pulse generator Ortec 419 relating the pulse heights provided by the reference Si detector with those from the diamond devices, in which an electron-hole pair creation energy of 13.2 eV was assumed [15]. The noise threshold was set to CCE= 0.08. The IBIC signals did not show any remarkable degradation in time, and significant polarization effects were not observed after ~$10^8$ protons/cm$^2$.

**Results and discussion** – The total CCE spectra acquired with the IBIC technique by the two diamond detectors polarized at -50V are reported in Fig. 2a, showing a significant enhancement in the charge collection performance and energy resolution of the 3D detector with respect to the planar structure, where the incomplete charge collection leads to a CCE below 50%.

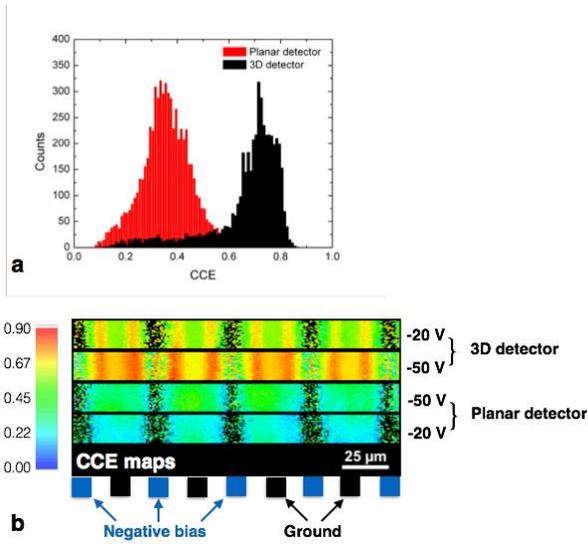

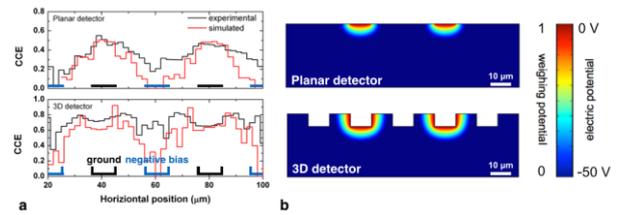

Fig. 2: (Colour on-line) a) Charge collection efficiency (CCE) spectra acquired by the 3D (black) and planar (red) detector at a negative bias of -50V. b) CCE maps acquired at a negative bias of -20V and -50V from by raster scanning a 1 MeV H+ micro-beam onto the surface of the 3D and planar detector. The position of the electrodes and their electrical connections is schematically depicted.

Fig. 2b shows IBIC maps collected when a voltage of -20 V and -50 V was applied to the two diamond detectors. The CCE is encoded in the color scale, which represents the median of the IBIC pulse distribution for each pixel. The IBIC map exhibits a periodic structure, where a high charge collection efficiency is achieved, alternated with a low CCE region. IBIC maps acquired at a positive bias configuration exhibited specular CCE features with respect to those reported in Fig. 2, i.e. the electrodes characterized by high and low CCE values were mutually inverted. Such behavior can be explained by considering the back-to-back geometry of the Schottky electrodes, which always results in the formation of a depletion region underneath the anodes, whose position depends on the bias polarity. According to the Shockley-Ramo-Gunn theory [25], the drift along the electric field lines of the excess charge carriers generated beneath the electrodes induces a charge at the sensitive electrode, thus producing detectable IBIC pulses. On the other hand, the diffusion of carriers in the neutral region of the device does not result in a measurable charge signal as long as they are not injected in the depletion region. As a consequence, the CCE maxima in Fig. 2b can be identified with the grounded interdigitated electrodes whereas a much smaller CCE contribution is observed around the other electrodes where a negative bias is applied. However, the active region of the detector, i.e. the region where non-zero IBIC pulses are collected, may extend beyond the actual size of the depletion region. In fact, a fraction of the minority carriers generated in the neutral region can be injected in the active region by diffusion, hence contributing to the induced current. In addition, the physical size of the ion micro-beam may lead to a local distortion of the median CCE value in the areas surrounding the depletion region. In fact, in areas where the charge collection efficiency is below the noise threshold level, the low amount of IBIC pulses, which are recorded from the surrounding regions due to the ion beam dispersion, may be sufficient to provide a significant contribution to the statistical analysis of the IBIC maps, thus leading to CCE values higher than expected in low efficiency regions.

Fig. 3: (Colour on-line) a) Experimental (black lines) and simulated (red lines) CCE profiles along the horizontal position extracted from the IBIC the maps in Fig. 3 at a bias of -50V for the planar (top view) and 3D (bottom view) detector; the relative position of the electrodes is schematically depicted. b) Electric potential and weighting potential distribution simulated for the planar detector (top view) and 3D detector (bottom view) at a negative bias of -50V assuming an effective acceptor concentration of $5\times10^{14}$ $cm^{-3}$ and $5\times10^{14}$ $cm^{-3}$, respectively.

As the applied bias voltage increases, the active region widens, allowing IBIC pulses to be detected even when ions hit the sample in the small gap (10 μm) separating two adjacent electrodes. The widening of the active regions is enhanced for the 3D detector due to the presence of the vertical electrodes, acting as planar back-to-back Schottky diodes. Therefore, the presence of the vertical electrodes in the 3D diamond detector further increases the charge collection efficiency indicating two additional CCE peaks at the edges of each grounded electrode.

In Fig. 3a the experimental CCE median profiles along a horizontal scan line of the IBIC maps (i.e. perpendicularly to the extension of the finger electrodes) are reported (black lines), together with the outline of the interdigitated electrodes, confirming the interpretation reported above.

In order to provide a comprehensive interpretation of the results of the IBIC experiments, a two-dimensional numerical finite-element model of the two devices under investigation was defined on the vertical cross-sections of the device. The electrostatic problem was evaluated by considering a 50 μm thick diamond layer. The electric potential was calculated as the numerical solution of the Poisson′s equation coupled with the relevant stationary drift-diffusion equation for hole and electron carriers, using the commercial Finite Element Method (FEM) software Comsol Multiphysics 4.3. Electron and hole mobilities in diamond were assumed to be 1700 and 2200 cm2 V-1 s-1, respectively [1, 26].

Based on reference values from diamond detectors with sandwich electrodes geometry grown with the same MPECVD technique, a Schottky barrier of 1.2 V was assumed [Ciancaglioni]. The possible effects on the internal electric field distribution due to the presence of charges trapped at the metal-diamond interface was assumed to be negligible due to the high applied biases considered in the present numerical simulations.

The color maps in Fig. 3b indicate the electric potential and the weighting potential distribution when a voltage of -50 V is applied to the biased electrodes. Numerical CCE profiles were then extracted according to the following procedure. The CCE maps as a function of electron-hole pair generation position were evaluated by the FEM solution of the charge carriers drift-diffusion adjoint equation [15]. Numerical CCE profiles were obtained by convoluting the relevant CCE maps with the Bragg′s generation profile of 1 MeV protons in diamond, calculated by the SRIM-2008 Monte Carlo code [24]. The resulting profile underwent a further processing, which took into account for associated experimental effects. Particularly, a CCE profile with a space step of 0.25 μm was determined considering the generation of 30 protons at each position. The ion beam size dispersion was simulated by choosing the effective position of incidence of each ion around the nominal value according to a Gaussian distribution of 7 μm FWHM. The single IBIC pulse corresponding to each position was then summed with normally distributed electronic noise (0.08 CCE) and Fano factor (0.08 CCE [28]). Finally, each ion pulse above the noise threshold level was used to evaluate a median CCE profile; a zero value was assigned when the number of ion pulses exceeding the noise threshold was smaller than 5. The obtained simulated CCE profiles are superimposed to the experimental data in Fig. 3a. The satisfactory agreement between experimental and simulated data can be appreciated assuming constant hole and electron lifetimes of $\tau_p$=1.3 ns and $\tau_n$=25 ps and effective concentration of $5\times10^{14}$ cm-3 and $7\times10^{14}$ cm$^{-3}$ acceptors for the 3D and planar detector, respectively. The carrier lifetimes and doping concentration were set as independent simulation parameters and were estimated on the basis of the reproduction of the experimental CCE profiles. The estimated lifetimes are consistent with previously reported works on namely identical diamond detectors[15,A]; the acceptor concentrations are within the typical variability range $(3-7)\times10^{14}$ cm$^{-3}$ reported for samples grown with the same MPECVD technique [21].

It is possible to make some final remarks on the shape of the CCE profiles for the two diamond detectors, allowing a deeper understanding of some of their characteristic features. For the planar detector, a significant decrease in the CCE profiles from the plateau above the grounded electrode occurs immediately outside its geometrical extension; the numerical results support the interpretation that the residual non-zero IBIC pulses outside the depletion region could be ascribed to the finite size of the ion micro-beam rather than to electron injection from the neutral region, due to their negligible contribution to the induced charge pulse formation. On the other hand, the CCE profile relevant to the 3D detector has two maxima in correspondence of the vertical electrodes and exhibits a decrease around the edges of the grounded electrodes, showing a wider active area. In both cases, the minimum of the CCE is at the centre of the negatively-biased electrodes, exhibiting a higher value for the 3D detector with respect to the planar detector.

The experimental plots reported in Fig. 4 show the behavior of the maxima of the CCE profiles as a function of the applied voltage. Both the experimental and the simulated data points report an uncertainty of 0.025 CCE taking into account for the statistical dispersion of the simulated IBIC pulses. It is worth noting that, for both the 3D and planar detector, CCE maxima acquired applying a positive bias at the read-out electrode display the same values with respect to the negatively-biased configurations analyzed in Fig. 2 and Fig. 3. The invariance of the absolute CCE values upon inversion of the applied bias confirms that polarization effects were negligible and therefore those charge trapping effects occurring at the metal-diamond interface could be neglected. The CCE of the 3D diamond detector increases as a function of the bias voltage up to a saturation value at ~0.85 CCE occurring at voltages higher than -40 V. On the contrary, the CCE of the planar detector increases up to ~0.55 CCE at -60 V without showing saturation effects. The incomplete charge collection exhibited by both detectors is ascribed to the partial depletion of the devices, since the electric field is negligible at depths below the grooves thickness (i.e. d > 6 μm); therefore, in this neutral region, the motion of the excess generated charge is limited by the short diffusion length resulting from the estimated carrier lifetimes. On the other hand, the different behavior of the planar and 3D detector is attributed to the different electrodes geometry. In the 3D detector, where the vertical electrodes create a parallel-plate diode, the carriers drift velocities reach their saturation velocities; on the other hand, the depleted volume in the planar detector widens at increasing bias and a full depletion is not achieved in the voltage range under consideration. The relevant CCE maxima obtained from the numerical models are superimposed in Fig.

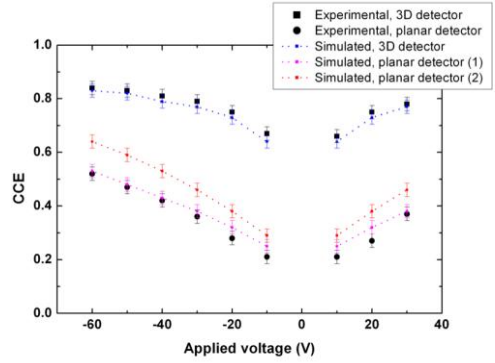

Fig. 4: (Colour on-line) CCE maxima extracted from the experimental profiles of the 3D (black squares) and planar (black circles) as a function of the voltage applied at the biased electrode. Data extracted from the numerical models are also shown. Blue and red scatters represent the CCE maxima of the 3D detector and planar profiles, respectively, simulated assuming an effective acceptor concentration $N_A$=$5\times10^{14}$ cm$^{-3}$. Purple scatters represent the CCE maxima of the planar detector simulated assuming $N_A$=$7\times10^{14}$ cm$^{-3}$

4 for both the 3D (blue line) and the planar (purple line) detector, showing an excellent agreement with the measured data. In order to provide a direct comparison of 3D and planar detectors properties, a numerical model of the CCE maxima for the planar detector was also performed using the same physical diamond parameters (i.e. effective acceptor concentration, carriers mobility, etc.) adopted for the 3D detector. The resulting curve is also reported in Fig. 4 (red line), and allows to appreciate, in any case, a significant enhancement of the 3D detector performance (up to ~30% of CCE increase at -50V) with respect to the planar geometry.

**Conclusions** − A CVD single-crystal diamond device with a novel three-dimensional interdigitated electrode geometry was successfully fabricated, modeled and characterized. Ion Beam Induced Charge (IBIC) measurements with 1 MeV H+ ions showed a significant increase in the detector's charge collection efficiency (CCE) performance and energy resolution if compared with standard surface metallic electrodes. In addition, the 3D diamond detector shows a wider active area if compared to the planar detector, as a result of the different geometry of the Schottky junctions in the two devices. The versatility of the micro-machining procedure allows in principle to tune the aspect ratio as well as the depth and spacing of the grooves, thus addressing the definition of optimal electrode geometries for the detection of specific ion energy and species. Moreover, it is possible to envisage that both surfaces of the device are microfabricated to increase extension of CCE-enhanced regions. Despite the fact that ions penetrating significantly deeper than the grooves would produce a similar response for the planar and 3D electrode devices, a significant enhancement of the induced charge signals is indeed expected when the ions stopping range is comparable with the depth extension of the grooves. Although an optimal electrode geometry (i.e. depth and aspect ratio of the grooves) is still under development, the observed significant performance improvements make this new architecture extremely appealing for the development of novel diamond-based 3D ionizing radiation detectors for applications in both nuclear science and dosimetry.

***

This activity was supported by the following projects, which are gratefully acknowledged: "DiaMed" project funded by the National Institute of Nuclear Physics (INFN); FIRB "Futuro in Ricerca 2010" project (CUP code: D11J11000450001) funded by the Italian Ministry for Teaching, University and Research (MIUR); "A.Di.N-Tech." project (CUP code: D15E13000130003) funded by University of Torino and Compagnia di San Paolo.